\documentclass{bmcart}

\usepackage[utf8]{inputenc}

\usepackage{graphicx}
\usepackage{amsbsy}

\startlocaldefs
\endlocaldefs

\begin{document}

\begin{frontmatter}

\begin{fmbox}
\dochead{Original article}

\title{Probing the transition from dislocation jamming to pinning by machine learning}

\author[
   addressref={aff1},                   
   corref={aff1},                       
   noteref={},                        
   email={henri.salmenjoki@aalto.fi}   
]{\inits{HS}\fnm{Henri} \snm{Salmenjoki}}
\author[
   addressref={aff2},
   email={lasse.laurson@tuni.fi}
]{\inits{LL}\fnm{Lasse} \snm{Laurson}}
\author[
   addressref={aff1,aff3},
   email={mikko.alava@aalto.fi}
]{\inits{MA}\fnm{Mikko J} \snm{Alava}}

\address[id=aff1]{
  \orgname{Aalto University, Department of Applied Physics}, 
  \street{PO Box 11000, 00076 Aalto},                     %
  \city{Espoo},                              
  \cny{Finland}                                    
}
\address[id=aff2]{%
  \orgname{Computational Physics Laboratory, Tampere University},
  \street{ P.O. Box 692, FI-33101},
  \city{Tampere},
  \cny{Finland}
}
\address[id=aff3]{%
  \orgname{NOMATEN Centre of Excellence, National Centre for Nuclear Research},
  \street{ A. Soltana 7, 05-400},
  \city{Otwock-Swierk},
  \cny{Poland}
}

\begin{artnotes}

\end{artnotes}

\end{fmbox}

\begin{abstractbox}

\begin{abstract} 
Collective motion of dislocations is governed by the obstacles they encounter. 
In pure crystals, dislocations form complex structures as they become jammed by their anisotropic shear stress fields. 
On the other hand, introducing disorder to the crystal causes dislocations to pin to these
impeding elements and, thus, leads to a competition between dislocation-dislocation and dislocation-disorder interactions.
Previous studies have shown that, depending on the dominating interaction, the mechanical response and the way the crystal yields change. 

Here we employ three-dimensional discrete dislocation dynamics simulations with varying density of fully coherent precipitates to study this phase transition $-$ from jamming to pinning $-$ using unsupervised machine learning.
By constructing descriptors characterizing the evolving dislocation configurations during constant loading, a confusion algorithm is shown to be able to distinguish the systems into two separate phases. These phases agree well with the observed changes in the relaxation rate during the loading. 
Our results also give insights on the structure of the dislocation networks in the two phases.

\end{abstract}

\begin{keyword}
\kwd{discrete dislocation dynamics}
\kwd{dislocation pinning}
\kwd{dislocation jamming}
\kwd{machine learning}
\end{keyword}

\end{abstractbox}

\end{frontmatter}

\section*{Introduction}
\label{intro}

While deforming, crystalline materials change irreversibly through discrete plastic events, i.e. avalanches, originating from the collective motion of dislocations -- the topological defects of the crystal lattice \cite{papanikolaou2017avalanches}.
These dislocation avalanches exhibit scale invariance with their distributions of sizes and durations encompassing several orders of magnitude \cite{zaiser2006scale}.
This has lead to the discussion of plastic deformation as a non-equilibrium phase transition: below critical loading, the dislocations merely jump from one configuration to another, and the actual yielding of the crystal occurs at the critical point of diverging avalanches and uninhibited flow of dislocations.
However, the dislocation movement has a highly complex nature arising from the interplay of the evolving, anisotropic interaction field produced by other dislocations and possible pinning field caused by disorder inside the crystal \cite{miguel2002dislocation,ardell1985precipitation}.
Thus, the collective dislocation behaviour is dictated by two competing phenomena $-$ dislocation-dislocation interaction induced \textit{jamming} and dislocation-obstacle induced \textit{pinning} $-$ that can be hard to distinguish from each other although they have fundamental differences \cite{ispanovity2014avalanches,sparks2018nontrivial,salmenjoki2019presiavals}.

Indeed in the case of dislocation jamming of pure dislocation systems, the interacting dislocations enter a state of 'extended criticality' where the system shows no distinct critical point but seems to recede in the constant vicinity of the transition independent of the loading force \cite{ispanovity2014avalanches,lehtinen2016glassy}. 
However, crystals are rarely completely pure, and introducing some disorder $-$ such as precipitates $-$ to the crystal to impede dislocation motion can increase the crystal's mechanical strength, and alter the criticality and ensuing avalanche behaviour of the system \cite{zhang2017taming,pan2019rotatable}. 
The key point here is that obstacles to dislocation motion may change the system behaviour by inducing dislocation pinning, which, if strong enough, results in a well-defined critical point of a depinning transition of the dislocation assembly \cite{ovaska2015quenched}.

Our recent study of 3D discrete dislocation dynamics (DDD) simulations of FCC aluminium with the inclusion of stationary fully coherent precipitates (see Fig.\ \ref{fig:-1}) showed that, by systematically increasing the strength or density of the precipitates, the system goes from the phase of dislocation-interaction dominated jamming to disorder-dominated pinning, and this transition can be observed in both constant load simulations as well as when quasistatically ramping up the external stress \cite{salmenjoki2019presiavals}. The related phenomenology depends on the loading protocol employed. For the quasistatic stress ramp simulations, one observes in general a sequence of strain bursts with a broad size distribution. In the jamming-dominated regime, the average strain burst size grows exponentially with the applied stress, while in the pinning phase we found a critical stress value where the average strain burst size exhibits a power-law divergence. Here, we
focus on the creep-like constant loading simulations with varying precipitate density $\rho_p$. There, the general behaviour in both of the phases, i.e. jamming and pinning, is on the one hand similar: In both phases the systems appear to possess a critical stress $\sigma_c(\rho_p)$ where one observes a power-law relaxation of the shear rate, $\dot{\varepsilon} \sim t^{-\theta}$. On the other hand, the relaxation becomes more rapid (larger $\theta$) as the systems move further into the pinning phase \cite{miguel2002dislocation,salmenjoki2019presiavals}.
This is illustrated in Fig.\ \ref{fig:0}.

An open question regarding the phase transition from jamming to pinning is how exactly does it alter the dislocation structures in the systems?
Furthermore, despite the apparent similarities in the response (i.e. power law relaxation) of the dislocation systems in the different phases, could one be able to distinguish them by their dislocation structures without specific {\it a priori} knowledge of the transition?
To address this problem, we use machine learning (ML). ML is proving to be a flexible and useful tool for physics and materials science \cite{zdeborova2017machine,mehta2019high,papanikolaou2018learning,papanikolaou2019spatial,steinberger2019machine,zhang2019extracting,yang2020learning}.
Using ML for the detection of phase transitions in statistical physics has given fruitful results \cite{carrasquilla2017machine,hu2017discovering,shirinyan2019self} and here we applied the unsupervised 'confusion' scheme introduced in \cite{van2017learning}. 
With the confusion algorithm, the only assumption one needs to make is that the system exhibits a phase transition in some control parameter range $-$ in our case, the control parameter being the precipitate density $\rho_p$ $-$  and the algorithm should be able to find the critical value $\rho_p^c$ by using the states of the systems as input.

Here we followed the evolving systems by concentrating on both the fine details and the long-ranged structures of the dislocation network.
To accomplish this we computed the dislocation junction lengths, geometrically necessary dislocation (GND) density and dislocation correlation, and used these separately to describe the microstructure for the ML algorithm.
Our results show that the algorithm was able to find the phase transition from all of the used descriptors and the discovered values of $\rho_p^c$ are in perfect agreement.
Therefore, as the dislocation structures in the two phases evolve in notably different ways, we were able to quantify some of the changes in the systems by analyzing the used structural descriptors. 
The rest of this paper is structured as follows:
the implementation of the ML method, along with the details of the DDD simulations and our approaches to characterize the dislocation structures, are presented in the next section. 
After the methodology, we proceed to show results obtained with the ML algorithm and we finish with some discussion.

\section*{Methods \label{sec:methods}}

\subsection*{DDD simulations}

We study the effect of varying the precipitate density $\rho_p$ on the nature of the collective dislocation dynamics within 3D DDD 
simulations using our modified version of the ParaDiS code \cite{arsenlis2007enabling,lehtinen2016multiscale}. 
ParaDiS implements the dislocation interactions by approximating the continuous dislocation lines by a set of straight dislocation segments.
The segments interact through the stress fields arising from the linear elasticity theory, while the diverging fields at dislocation cores are replaced by the use of results from molecular dynamics simulations. 
To cope with the long-range elastic forces, ParaDiS uses multipole expansion. 
Our version of ParaDiS also enables including disorder to the system, in the form of spherical precipitates \cite{lehtinen2016multiscale}.
The precipitates are frozen pinning sites for the dislocations that produce a short-range radial force
\begin{equation}
    F(r) = \frac{2 A r e^{-r^2/r_p^2}}{r_p^2},
\end{equation}
where $A$ is a constant, $r$ is the distance from the precipitate to the dislocation and $r_p$ is the radius of the precipitate. 
In the context of transition from dislocation-dominated jamming to disorder-dominated pinning, the relevant parameters are the precipitate density $\rho_p$ and the precipitate strength $A$ \cite{salmenjoki2019presiavals}. 

For our simulations, we set parameters to approximate those of FCC aluminium with precipitates of fixed strength and size in a simulation box with periodic boundaries $-$ $A$ was especially chosen so that the system exhibits both jamming and pinning-dominated response depending on $\rho_p$ \cite{salmenjoki2019presiavals}. 
The parameters are presented in Table \ref{table:simparams}.

The simulations started with two relaxation periods, the first with only the dislocation networks and the second with also the precipitates present, to ensure the systems reached meta-stable states.
After the initial relaxation, the systems were driven by applying a constant external stress $\sigma$.
Depending on the magnitude of driving force, the systems tend to either get stuck (exponential decay of strain rate $\dot{\varepsilon}$ with small $\sigma$) or reach linear creep-like conditions (constant $\dot{\varepsilon}$ with large $\sigma$).
However independent of precipitate density, all of the systems possess also a critical value $\sigma_c$ (dependent on $\rho_p$, see Table \ref{table:sigmac}) that leads to a power-law relaxation of $\dot{\varepsilon}$, as seen in Fig.\, \ref{fig:0} \cite{miguel2002dislocation,salmenjoki2019presiavals}. 
The effect of precipitate density is seen in the rate and starting time of the power-law decay: there is a transition between the behaviour of less disordered systems with $\dot{\varepsilon} \sim t^{-0.3}$ and more disordered systems with the more rapid decay starting earlier. 
To see how this transition affects the system, we characterized the dislocation structure and observed its evolution during the constant stress loading with $\sigma=\sigma_c$.

\subsection*{Characterizing dislocation structures}

In the characterization of the dislocation structures, we used three distinct descriptors. 
First,  we exploited the fact that disorder causes the dislocations to stretch when parts of the dislocations get pinned. 
With this in mind, we measured the length of dislocation links between two junction nodes \cite{sills2018dislocation} along the dislocation segments $l_{\mathrm{along}}$, and compared this to the shortest possible length between the nodes $l_{\mathrm{shortest}}$.
Thus, we define parameter $J$, 
\begin{equation}
J = l_{\mathrm{along}} - l_{\mathrm{shortest}},
\label{eq:j}
\end{equation}
which represents the roughness of a dislocation and by collecting its distribution inside a system provides information on the dislocation structure. 
As an example, Fig.\, \ref{fig:1}a shows the distribution of $J$ in the simulated systems. 

The second used descriptor was GND density \cite{arsenlis1999crystallographic,steinberger2019machine}. 
We computed the local GND density (the total GND density is constant throughout the simulation \cite{bulatov2000periodic}) by first evaluating the Nye tensor $\boldsymbol{\alpha}$ in voxels by
\begin{equation}
\boldsymbol{\alpha} = \frac{1}{V_{voxel}} \sum_i  \mathbf{b}_i \otimes \mathbf{l}_i,
\end{equation}
where $V_{voxel}$ is the voxel volume, $\mathbf{b}$ is the Burgers vector, $\mathbf{l}$ is the line direction giving also the segment length and the sum is over all dislocation segments $i$ inside the voxel. 
Then, the GND density $\rho_{GND}$ was  calculated from the Nye tensor. 
The resulting GND density fields, for instance the one with $10\times 10 \times 10$ voxels illustrated in Fig.\, \ref{fig:1}b, are quite system specific, and as we are interested especially in the changes in the dislocation structure, we focused on the evolution of GND density, i.e. $\rho_{GND}'(t) = \rho_{GND}(t) - \rho_{GND}(0)$. 
Moreover to remove the effect of periodic boundaries, we took the Fourier transform of $\rho_{GND}'(t)$ as we collected the data.

As the third and final descriptor, we calculated the dislocation spacing correlation according to \cite{csikor2007range}
\begin{equation}
C(r) = \left( \frac{\mathrm{d}}{\mathrm{d}r} L(r)  \right) / (4 \pi r^2 \rho),
\label{eq:correlation}
\end{equation}
where $\rho$ is the total dislocation density and $L(r)$ is approximated by computing the mean line length in spheres of radius $r$ centered at random points along the dislocation structure. 
We note that in the case of 2D DDD simulations, the average dislocation-dislocation correlation function changed drastically when mobile solutes (pinning points) were introduced to the system \cite{ovaska2016collective}. 
Here we focused on longer-range correlations to avoid possible effects caused by the assigned segment length restrictions of the computations. 
Fig.\, \ref{fig:1}c shows the dislocation correlation in systems with varying $\rho_p$. 

We proceed by collecting the descriptors listed above during the loading with $\sigma=\sigma_c(\rho_p)$ at intervals of $t = 10^{-9}\, \mathrm{s} = 2.6 \cdot 10^{5}\, G M$, where the times are given in the units of shear modulus $G$ times the dislocation mobility $M = M_{\mathrm{edge}} = M_{\mathrm{screw}}$. 
Due to computational challenges of 3D DDD simulations, we simulated only 19 systems for every value of $\rho_p$ and $\sigma_c$. The time reached in every simulation was at least $4.7 \cdot 10^{6}\, GM$ although some systems were able to run even longer in their allocated simulation time.

\subsection*{Unsupervised learning of the phase transition}

To observe the transition from dislocation jamming to pinning in an unsupervised manner, we used the confusion method presented in \cite{van2017learning}. 
The idea is that, assuming the studied system experiences a transition in a control parameter range (in our case $[\rho_p^0, \rho_p^1]$) with some value $\rho_p^c$, one expects that the different systems below and above $\rho_p^c$ are distinguishable from each other. 
Thus by appointing trial values $\rho_p'$ in the range $[\rho_p^0, \rho_p^1]$, the sample systems are assigned to classes depending on whether $\rho_p$ is below or above $\rho_p'$. 
This way, a machine learning classifier trained on the trial samples in supervised fashion should perform best near the critical point $\rho_p' \approx \rho_p^c$ where the systems are truly distinguishable.
Correspondingly further from $\rho_p^c$, the classifications should get worse as some of the samples are wrongly labeled.
If for instance a system was in jamming state (with $\rho_p<\rho_p^c$), trial value $\rho_p'<\rho_p$ would lead to the system being mislabeled to the pinning state with samples that actually belong to the pinning state. Then this labeling would be especially challenging for the classifier to learn because some of the samples in jamming state should be classified as jamming but some as pinning  - therefore the confusion. 
Observing the accuracy of the classification in the range $[\rho_p^0, \rho_p^1]$ should therefore be somewhat $W$-shaped, as the accuracy is good at the transition but also at the beginning and at the end of the range (as large majority of the samples are labeled to one class, the classifier gets high score by simply predicting always the majority class).

As we were dealing with a data set of 190 systems with more than one thousand of collected features, we applied some dimensionality reduction before teaching any classifier. 
The three distinct data sets (different descriptors) were cast to lower dimensions by principal component analysis (PCA). 
In PCA, every feature of the data is first scaled to zero mean and unit variance, and then the entire dataset is represented by $n$ orthogonal linear combinations of the original data which maximize amount of explained variance. 
This happens in descending order, so that with the first principal component (PC), the explained variance is the largest. 
Fig.\, \ref{fig:2} already shows that by  projecting the data to the space of the two first PCs, there is a rather smooth transition in dislocation structures from less to more disordered landscapes with all of the used descriptors.

\subsection*{Supervised classifier for the confusion method}

For a classifier, our choice was based on linear discriminant analysis (LDA) \cite{bishop2006pattern}. 
In this simple case of 2-class classification, LDA builds one linear decision boundary into the input space according to 
\begin{equation}
    y (\mathbf{x}) = \mathbf{w}^T \mathbf{x} + w_0,
    \label{eq:boundary}
\end{equation}
where $\mathbf{x}$ is the feature vector of a sample, $\mathbf{w}$ and $w_0$ are the weights and bias of the classifier and $y(\mathbf{x}) = 0$ is the boundary. 
We used the implementation by \textit{scikit-learn} \cite{scikit-learn}, that computes the boundary parameters by assuming that the samples inside different classes are Gaussian distributed, i.e. probability of a sample with features $\mathbf{x}$ when belonging to class $k$ is 
\begin{equation}
    P(\mathbf{x} | y=k) = \frac{1}{ (2 \pi)^{d/2} | \Sigma_k |^{1/2}} \exp \left( -\frac{1}{2} (\mathbf{x} - \mu_k)^T \Sigma_k^{-1} (\mathbf{x} - \mu_k) \right)
    \label{eq:gaussian}
\end{equation}
 where $d$ is the length of $\mathbf{x}$, $\mu_k$ is the class-specific mean of features and $\Sigma_k$ is the covariance matrix.
Moreover to obtain a linear boundary, the different classes are assumed to have identical covariance matrices, so in our case of two classes, $k=-1$ or $k=1$, $\Sigma_{-1}  = \Sigma_1 = \Sigma$.
The weights for the decision boundary are obtained by applying the Bayes theorem, as at the boundary the probabilities of different classes given the sample are equal $P(y=-1|\mathbf{x}) = P(y=1 |\mathbf{x})$ and, thus, the log-probability ratio is 
\begin{equation}
    \log \left( \frac{P(y=1 |\mathbf{x})}{P(y=-1 |\mathbf{x})} \right) = \log \left( \frac{P(\mathbf{x}| y=1) P (y=1)}{P(\mathbf{x} | y=-1) P (y=-1)} \right) = 0.
\end{equation}
From this, the final weights,  $\mathbf{w}$ and $w_0$,  are obtained by substituting the probability distribution of Eq. \ref{eq:gaussian} and comparing to Eq. \ref{eq:boundary}, 
\begin{equation}
(\mu_{1} - \mu_{-1})^T \Sigma^{-1} \mathbf{x} - \frac{1}{2} (\mu_{1}^T \Sigma^{-1} \mu_{1} - \mu_{-1}^T \Sigma^{-1} \mu_{-1}) + \log \left( \frac{P(y=1)}{P(y=-1)} \right) = 0
\end{equation}
The LDA classifiers were evaluated by the straightforward accuracy, i.e. score $S = $ number of correctly predicted test samples $/$ number of test samples, and trained by 2-fold cross-validation to provide some tentative  confidence intervals.

\section*{Results \label{sec:results}}

The confusion curves with the different dislocation structure descriptors in Fig.\ \ref{fig:3}a show the expected $W$-shape.
What is striking, is that every curve shows a possible transition in the form of local maximum at the same spot, $\rho_p^c \approx 3\cdot 10^{19}\, \mathrm{m}^{-3}$.
Moreover, the classifying accuracy there is extremely good as every descriptor achieved score larger than $0.95$ at the local maximum. 
Comparing the position of the transition to the relaxation curves with different $\rho_p$ of Fig.\, \ref{fig:0}b and their power-law part represented by the exponents $\theta$ presented in Fig.\ \ref{fig:3}b, we see that the relaxation behaviour is distributed nicely to the two phases so that $\theta$ is close to constant on one side (jamming) of the transition, while on the other side it starts to increase (pinning) \cite{salmenjoki2019presiavals}.
Of course with $\rho_p = 5.1 \cdot 10^{19}\, \mathrm{m}^{-3}$ $\theta$ seems to still be close to the constant value of the jamming-side of the transition, but there the error of $\theta$, arising from the fact that the $\sigma_c$ used in simulations is impossible to get spot on to the one producing power-law relaxation, is notably higher than with other $\rho_p$.

The used number of PCs for the best confusion curves (i.e. the curve with the highest maximum accuracy somewhere else than the ends of the range) was 5 for junctions and GND density, and 10 for correlations.
Interestingly, the confusion curve obtained with junction lengthening data shows another distinct maximum near $\rho_p \approx 3 \cdot 10^{20}\, \mathrm{m}^{-3}$, although there the accuracy is not as good as at $\rho_p^c$. 
Similar fluctuations from the pure $W$-shape are also observed in Fig.\, \ref{fig:4} which shows the confusion curves with different amount of PCs used for the classifying task.
Basically all of the secondary maxima are positioned to the more disordered side with $\rho_p > \rho_p^c$.
Most likely this arises from the fact that in the pinning phase, the systems get more and more pinned with growing $\rho_p$ yielding faster relaxation with larger $\theta$ causing these systems to possess some distinguishability from each other despite being in the same phase.
This also explains the tendency of slightly asymmetric $W$-shaped curves in Fig. \ref{fig:4}, as the LDA score does not drop as much in the pinning phase as in the jamming phase.
But as was ensured by the choice of the best confusion curves, the dominant maximum is indeed near $\rho_p^c$.

We can also study how the ability of the confusion scheme to distinguish the two phases using different microstructure descriptors evolves in time by computing the confusion curves based on single snapshot structures, presented in Fig.\, \ref{fig:5}. 
There the classifiers were trained by using two PCs of the dislocation structure at the specific times.
Starting from the junction lengthening in Fig.\, \ref{fig:5}a, there seems to be a short transient time until the single time step curves have converged to close to the shape of the best confusion curve in Fig.\, \ref{fig:3}. 
This indicates that the junction lengthening shows early on the signs of distinct jamming and pinning phases. 
On the other hand GND density in Fig.\, \ref{fig:5}b, which was measured as the difference to the initial density field, shows that the phases are separated well in the immediate beginning of the driving.
However, the information about the transition is lost if looking at a momentary GND density field compared to one before loading.
Trying the confusion scheme to GND density field without extracting the initial field or difference in the field of subsequent time steps yielded no observable phase transition (not shown here).
Finally with the observed dislocation correlation functions in Fig.\, \ref{fig:5}c, the behaviour is similar as with junction lengthening: There is now a longer time during which the transition is not observed, but after that the curves start to resemble the best confusion curve with maximum at $\rho_p^c$. 
Again, this is quite evident because the correlation functions focused on long-range structures, so it takes time until the systems have evolved structures that are noticeably different in the two phases.
Notable here is also that the converged confusion curves are quite flat in the pinning phase.

\section*{Discussion \label{sec:conlusions}}

As the results with the unsupervised ML scheme showed, the dislocation configurations can be separated into two phases with different relaxation rates even though the general response, i.e. the power-law relaxation, is similar in the two cases.
The confusion scheme succeeded extremely well, as it was able to achieve accuracy $>0.95$ at the observed transition indicating that the systems where the dislocation-dislocation interactions dominate are significantly different from the precipitate-dominated systems.
This was further supported by the fact that all the three dislocation structure characterization metrics considered captured the transition happening at the same value of $\rho_p^c$ where also the relaxation starts to turn more rapid.

The success of all three descriptors reveals some of the notable differences between dislocation structures in the two phases. 
Firstly, the distribution of the junction lengthening $J$ captures the bowing of the dislocation lines and, clearly, the pinning points cause more stretching and bowing of junctions than the other possible obstacles, namely the jamming dislocation structures, as depicted already in Fig.\, \ref{fig:1}a.
Secondly, the spacing correlation of the dislocations, $C(r)$ shows that even long-range structures are slightly affected, although there the differences seem to arise more from the magnitude (and scaling by the total dislocation density in Eq. \ref{eq:correlation}) than the shape of the correlation functions which are plotted in Fig.\, \ref{fig:1}c.

Thirdly, the evolution of the local GND density finds similar structural changes as the other two descriptors: On one hand, the bowing dislocations are seen as a 'spreading' density of GND, while on the other hand with only few precipitates dislocations tend to move more in their straight forms. 
This is illustrated in Fig. \ref{fig:6} which shows the probability of a computational voxel having a non-zero GND density as a function of simulation time for different $\rho_p$. 
The systems in the jamming phase show more or less constant number of active voxels as dislocations keep their shape while in the pinning phase the number is clearly increasing as dislocations bow.
This happens despite the fact that the total GND density stays constant during the simulations. 
Undoubtedly, the effectiveness of GND density as a descriptor of the phase transition is also enhanced by the fact that in the pinning phase $\sigma_c$ is larger (faster changes in the dislocation structure right in the start of the simulation) but relaxation is more rapid (more constant structures on longer time-scale). 
However as Fig.\ \ref{fig:7} shows, the confusion scheme seems to be quite robust with respect to the resolution of GND density computation: even sparse number of voxels reveals the changes in the evolving structures.

To conclude our findings, we have studied the transition between dislocation-dislocation interaction dominated jamming and disorder dominated pinning. 
By tuning the disorder content through precipitate density and strength, the system changes the mechanical response and yielding  which is also seen in the power-law relaxation rate during the plastic flow with constant loading.
Here we have been able to distinguish the simulated systems to the two phases of jamming and pinning solely by their dislocation structures during the constant stress simulations and, thus, highlighted the changes in the microstructure caused by the phase transition. 
These results offer two obvious prospects for future study: first, to conduct further simulations of the borderline case system where neither dislocation-dislocation nor dislocation-precipitate interaction dominates over the other. The second one is that our results tell that the dislocation 
structures are different in the two phases. This means that one can correlate
these with the most interesting engineering quantity, the yield strength, possibly on a sample-to-sample basis as well. One should thus use the dislocation structure -oriented approach in the experimental verification of the different phases of crystal plasticity and for strength prediction.

\begin{backmatter}

\section*{    Availability of data and materials}
The data that support the findings of this study are available from the corresponding authors on reasonable request.

\section*{Competing interests}
  The authors declare that they have no competing interests.

\section*{    Funding}
The authors acknowledge support from the Academy of Finland Center of Excellence program, 278367.
LL acknowledges the support of the Academy of Finland via the Academy Project COPLAST (project no. 322405), 
and HS acknowledges the support from Finnish Foundation for Technology Promotion. 
MA acknowledges support from the European Union Horizon 2020 research and innovation
programme under grant agreement No 857470 and from European Regional Development Fund via Foundation for Polish Science International Research Agenda PLUS programme grant No MAB PLUS/2018/8.

\section*{Author's contributions}
HS, LL and MA designed the study. HS performed the simulations and data analysis, and wrote the first version of the manuscript. All authors contributed to the final version of the manuscript. 

\section*{Acknowledgements}
The authors acknowledge the computational resources provided by the Aalto University 
School of Science ``Science-IT'' project, as well as those provided by CSC (Finland).

\bibliographystyle{bmc-mathphys} 
\bibliography{refs}      

\section*{Figures}

\begin{figure}[h!]
\centering
\includegraphics[width=0.55\linewidth]{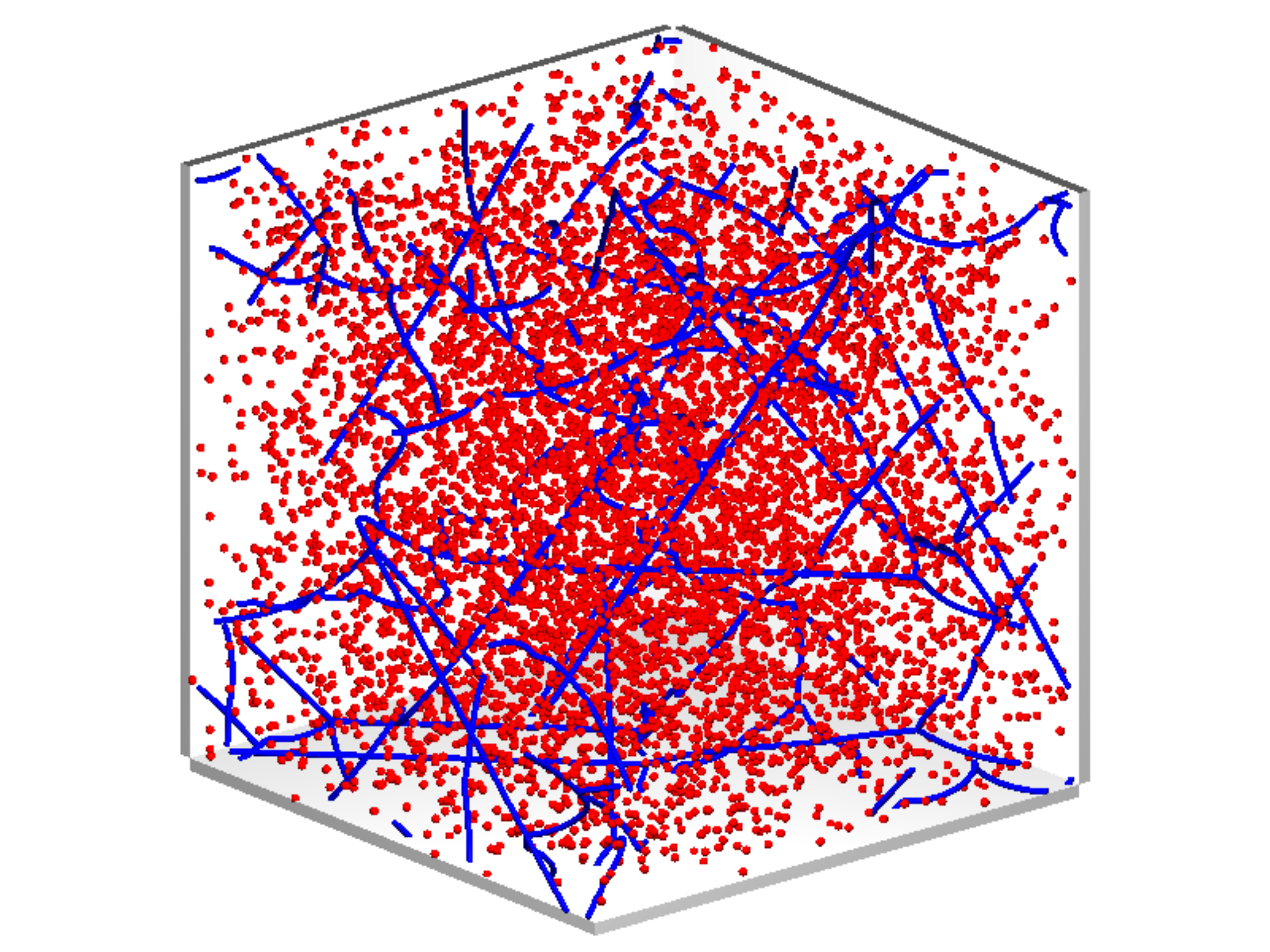}
\caption{ \csentence{ Snapshot of a simulated system.} The parameters were set to  $\rho_p = 10^{20}\, \mathrm{m}^{-3}$ and the image is taken at $t = 10^{-9}\, \mathrm{s} = 2.6 \cdot 10^{5}\, G M$, where $M=M_{\mathrm{edge}} = M_{\mathrm{screw}}$.}
\label{fig:-1}   
\end{figure}

\begin{figure}[h!]
\centering
\includegraphics[width=0.55\linewidth]{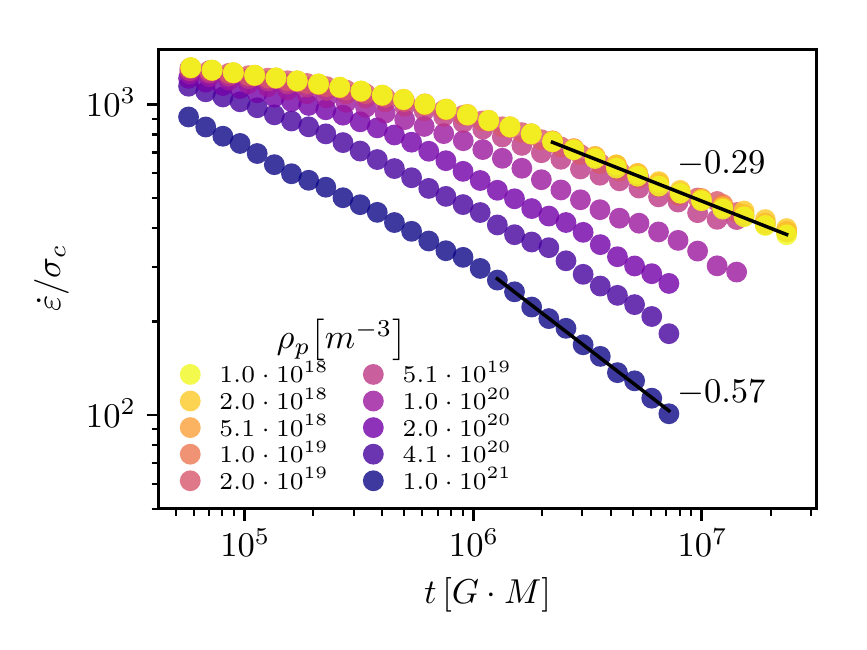}
\caption{ \csentence{Simulations of power-law relaxation  of dislocation systems with varying $\rho_p$.}  The figure shows the average strain rate during constant loading with $\sigma_c (\rho_p)$. The relaxation becomes more rapid and the transient time before power-law decay of $\dot{\varepsilon}$ decreases as $\rho_p$ is increased above some threshold value of $\rho_p$.    }
\label{fig:0}   
\end{figure}

\begin{figure}[h!]
\centering
\includegraphics[width=0.55\linewidth]{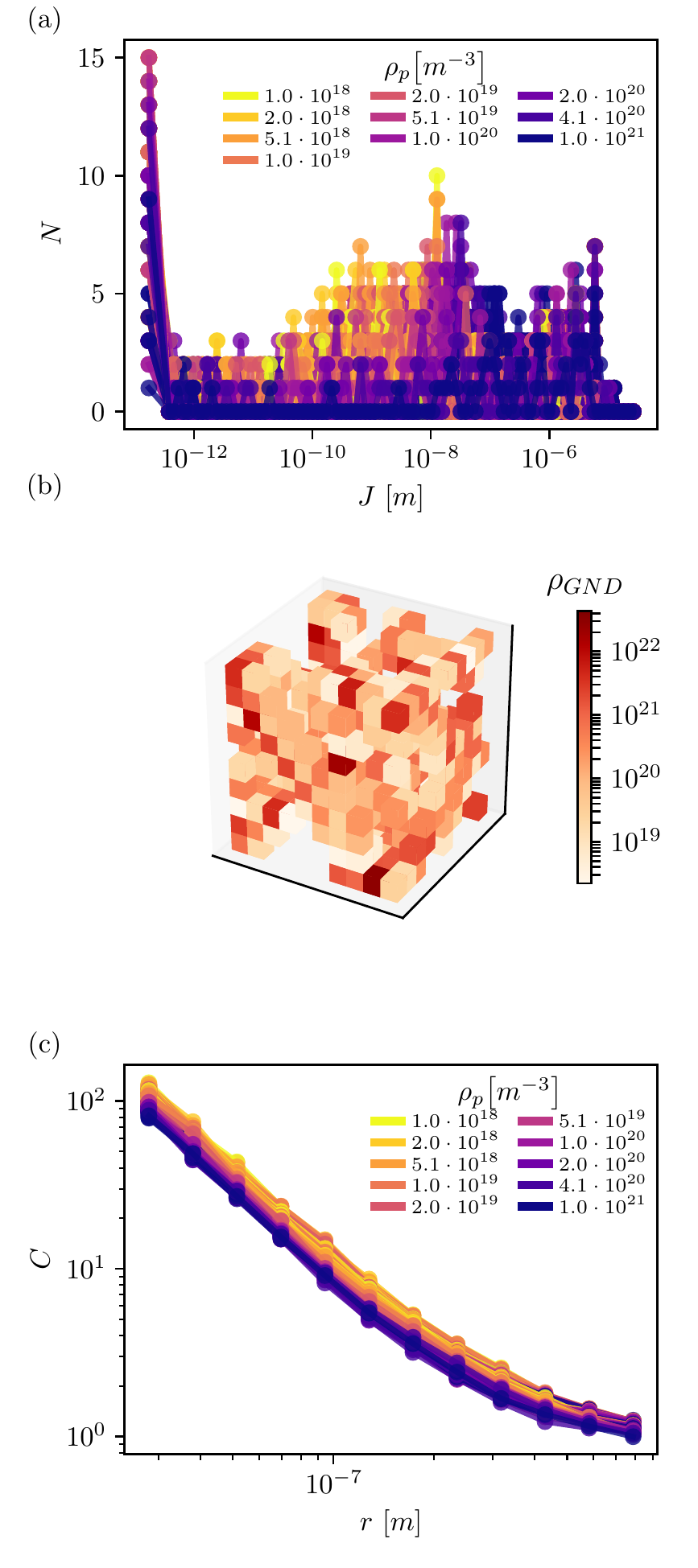}
\caption{ \csentence{Different ways to measure the dislocation structure.} \textit{(a)} Distribution of dislocation junction lengthening $J$ (Eq. \ref{eq:j}) in single systems  at $t = 2.6 \cdot 10^{5}\, G M$. The first bin also includes junctions with $J=0$.   \textit{(b)} An example of internal GND density  at $t = 2.6 \cdot 10^{5}\, G M$ in a system with $\rho_p = 10^{20}\, \mathrm{m}^{-3}$.  \textit{(c)} Dislocation  spacing correlation according to Eq. \ref{eq:correlation} in single systems at  $t = 5.2 \cdot 10^{6}\, G M$.   }
\label{fig:1}   
\end{figure}

  \begin{figure}[h!]
  \centering
  \includegraphics[width=0.55\linewidth]{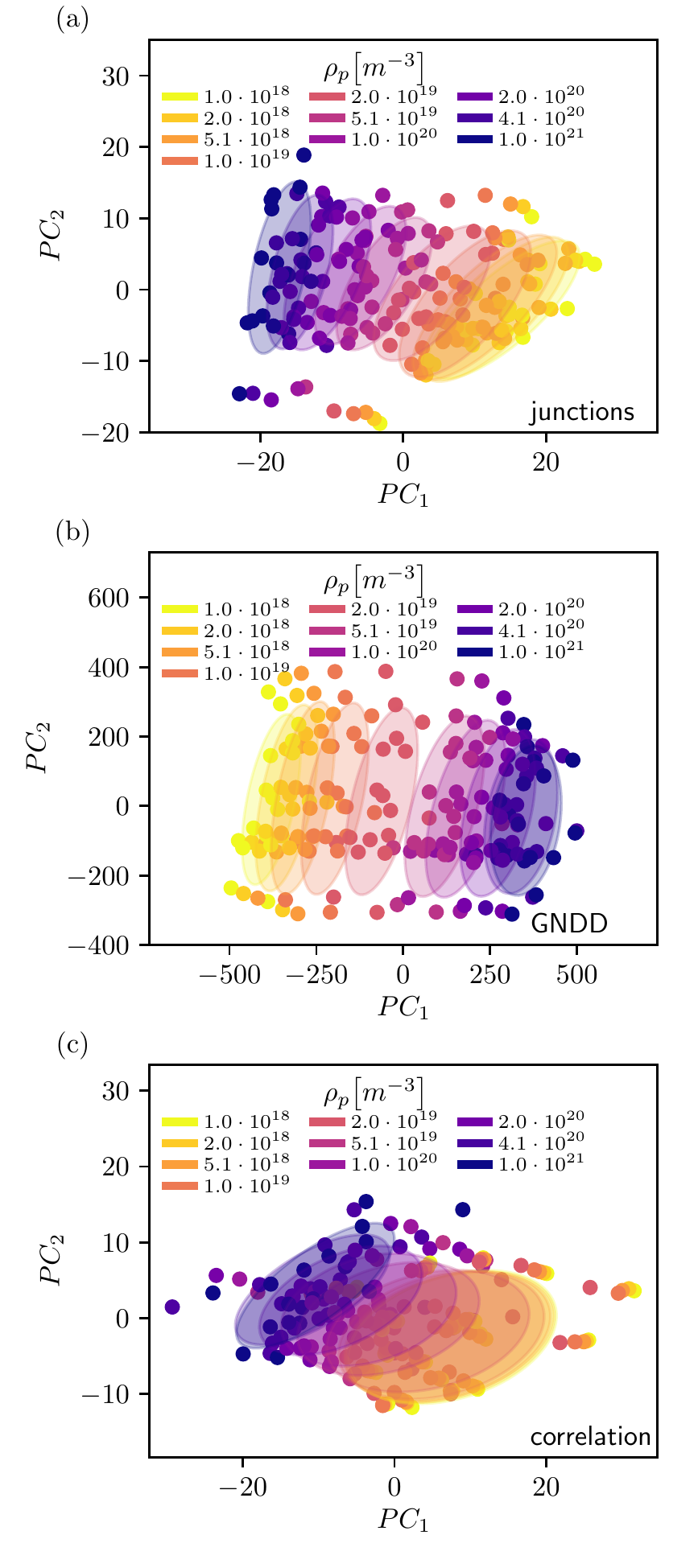}
\caption{ \csentence{The systems represented by the two first principal components.} The descriptors are \textit{(a)} junction lengthening \textit{(b)} GND density change and \textit{(c)} dislocation correlation collected during the simulation time interval. The coloured regions come from a fit of bivariate normal distribution on the data sets of different $\rho_p$ to visualize the differences between them.  } 
\label{fig:2}  
      \end{figure}

  \begin{figure}[h!]
\centering
\includegraphics[width=0.55\linewidth]{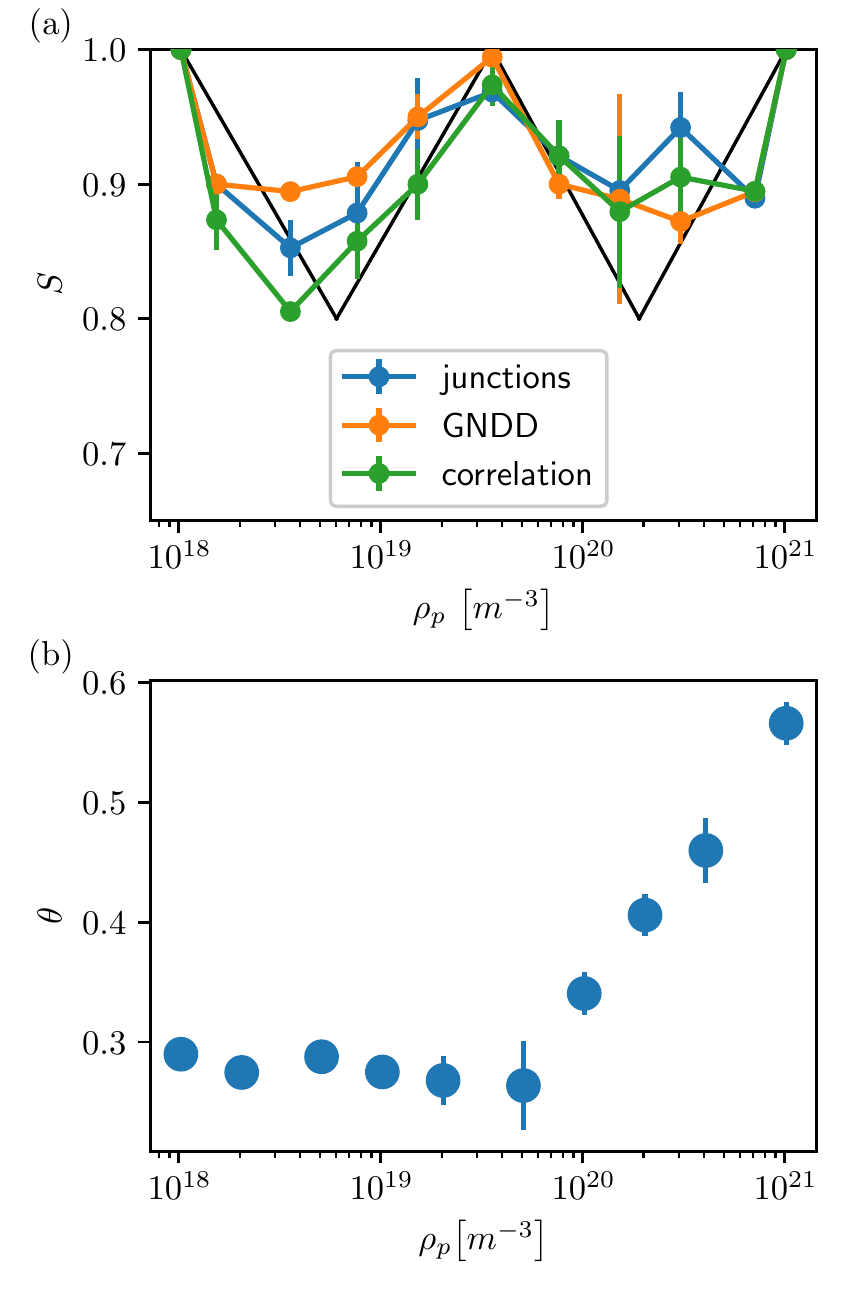}
\caption{ \csentence{The confusion curves compared to the exponent of relaxation} \textit{(a)} The confusion curves (classifier accuracy as a function of chosen threshold $\rho_p$) of different dislocation structure descriptors reach a maximum at the same $\rho_p$. For these curves, the used number of PCs was 5 for junctions and GND density, and 10 for correlations.  The error bars represent the standard deviation of classifier accuracy in 2-fold cross validation.  \textit{(b)} The maximum accuracy coincides quite well with the change in the relaxation rate,  depicted by the exponent $\theta$ of power-law part in the strain rate curves of Fig.\ \ref{fig:0}.  }
\label{fig:3}   
      \end{figure}

  \begin{figure}[h!]
\centering
\includegraphics[width=0.55\linewidth]{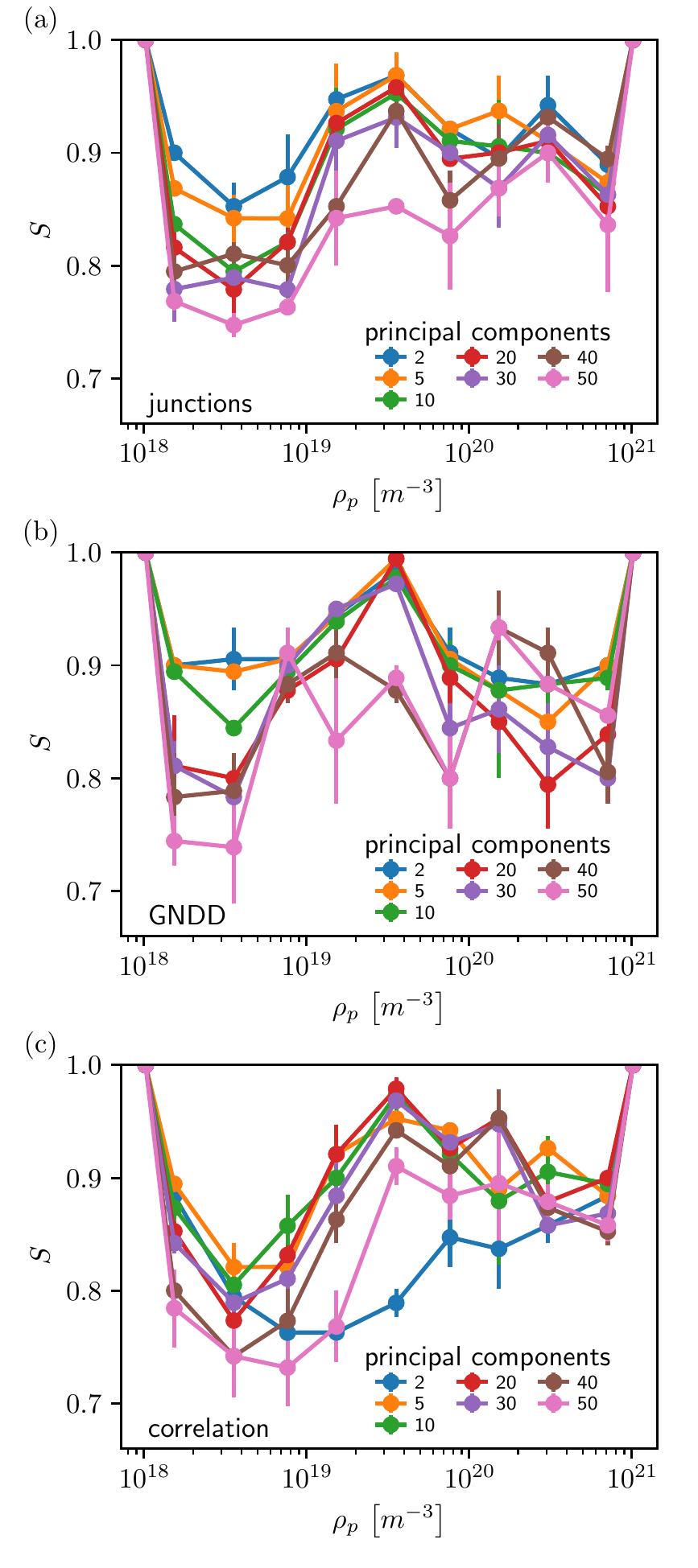}
\caption{ \csentence{The confusion curves with different number of principal components.} Used features are  \textit{(a)} junction lengthening, \textit{(b)} GND density change, and \textit{(c)} dislocation correlation during the simulation time interval. } 
\label{fig:4}   
      \end{figure}

  \begin{figure}[h!]
   \centering
\includegraphics[width=0.55\linewidth]{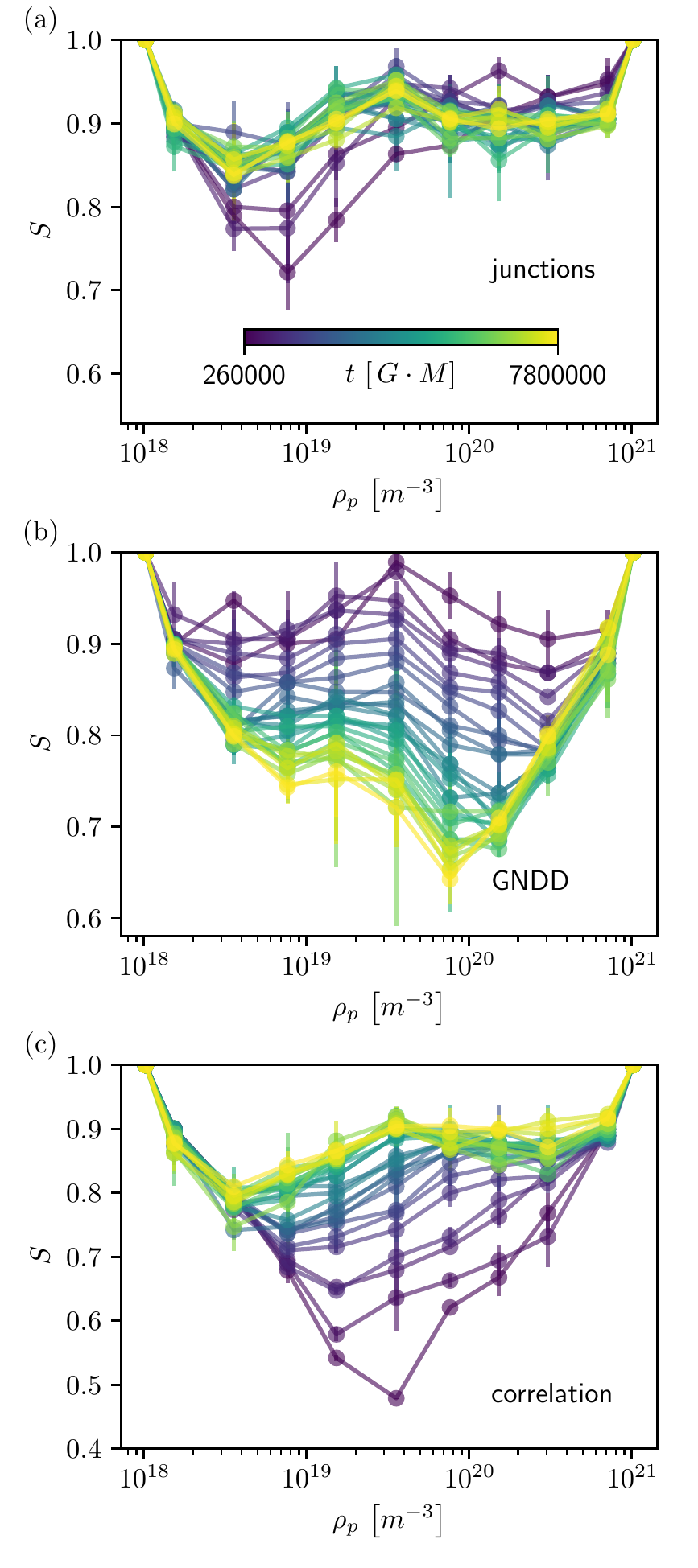}
\caption{\csentence{The confusion curves with two principal components of single time step snapshot.} Used features are \textit{(a)} junction lengthening, \textit{(b)} GND density change, \textit{(c)} dislocation correlation. The number of collected systems starts to decrease after $t= 4.7 \cdot 10^{6}\, G M$. }  
\label{fig:5}
      \end{figure}

  \begin{figure}[h!]
    \centering
\includegraphics[width=0.55\linewidth]{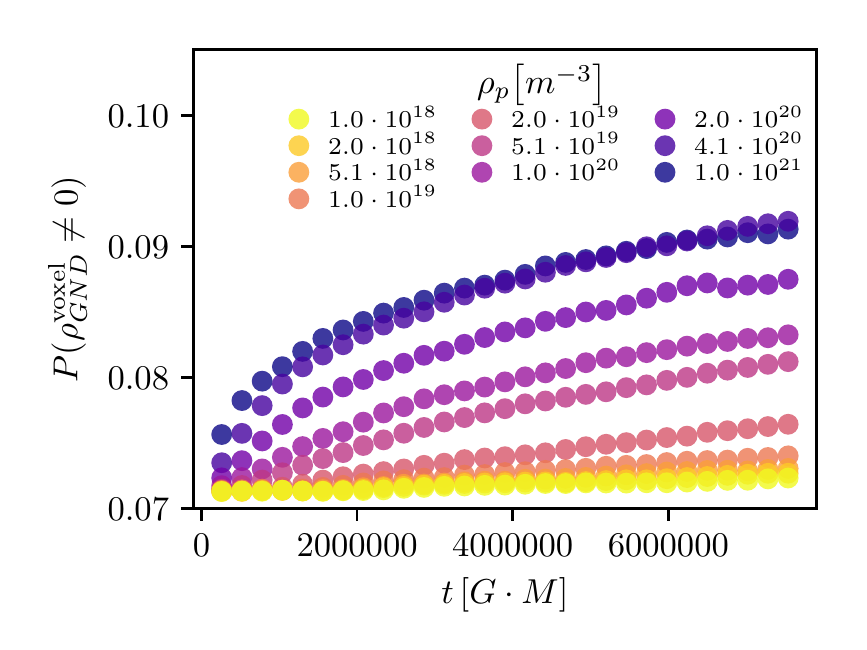}
\caption{ \csentence{The probability of finding the GND density inside a voxel to be non-zero during the simulations.} The results averaged over systems with specific $\rho_p$. The number of voxels used here was $25 \times 25 \times 25$, but similar observations were made with other number of voxels as well.  }
\label{fig:6} 
      \end{figure}

  \begin{figure}[h!]
    \centering
  \includegraphics[width=0.55\linewidth]{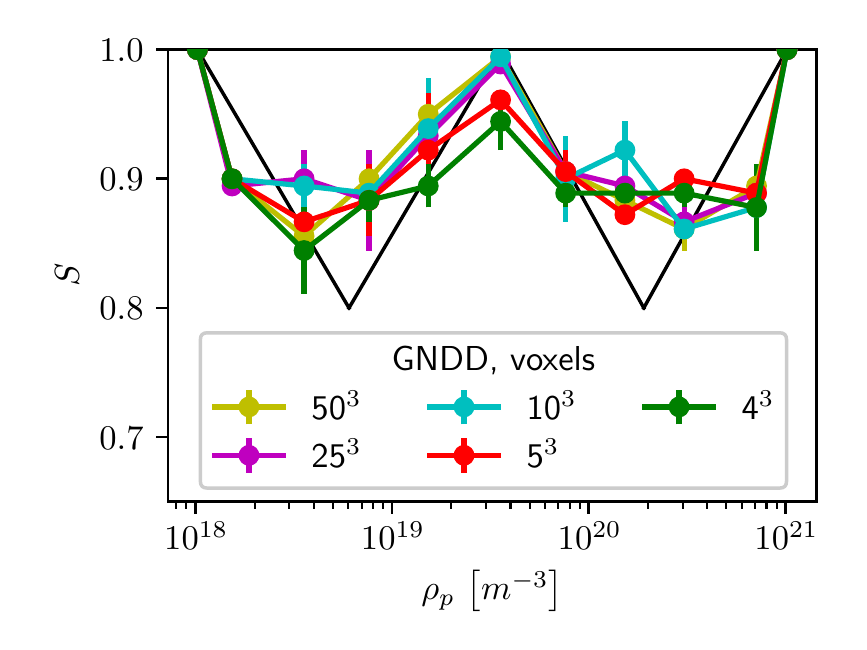}
\caption{ \csentence{Confusion curve dependency on the number of GND density voxels.}  Only slight decrease of the maximum in the curve is observed depending on the number of voxels (i.e. voxel size from $280 b$ to $3500 b$) when measuring GND density. }
\label{fig:7}    
      \end{figure}

 \section*{Tables}

\begin{table}
\caption{Simulation parameters.\label{table:simparams} }
\begin{center}
\begin{tabular}{ll}
Parameter	& Value   \\ \hline
System size $L$ & $4\, \mu\mathrm{m}$\\
Initial dislocation density $\rho_{0}$& $2.0\times 10^{12}\,\mathrm{m}^{-2} $ \\
$b$	& $0.286\,\mathrm{nm} $     \\
$r_{\mathrm{core}}$	& $5.0 \,b $     \\
Maximum segment length	& $80.0 \,b $     \\
$G$	& $26\, \mathrm{GPa}$     \\
$\nu$	& $0.35$     \\
$T$	& $300\, \mathrm{K}$     \\
ParaDiS mobility function		&$\mathrm{FCC}\_{0b}$				\\
$M_{\mathrm{edge}}$		&$10000.0\, (\mathrm{Pa}\,\mathrm{s})^{-1}$				\\
$M_{\mathrm{screw}}$		&$10000.0\, (\mathrm{Pa}\,\mathrm{s})^{-1}$				\\
$A$                     & $2.3 \cdot 10^{-19}\,\mathrm{Pa}\,\mathrm{m}^3$ \\
$r_{p}$ & $28.6 \,\mathrm{nm} $ \\
\end{tabular}
\end{center}
\end{table}

\begin{table}
\caption{The used critical values of $\sigma_c$ with different values of $\rho_p$. \label{table:sigmac} }
\begin{center}
\begin{tabular}{|c|c|}

$\rho_p\, \left[ \mathrm{m}^{-3} \right]$ & $\sigma_c \left[ \cdot 10^7\, \mathrm{Pa}\right]$\\
\hline
$1.0\cdot 10^{18}$ & $1.10$ \\  
$2.0\cdot 10^{18}$ & $1.25$ \\  
$5.1\cdot 10^{18}$ & $1.40$ \\  
$1.0\cdot 10^{19}$ & $1.70$ \\  
$2.0\cdot 10^{19}$ & $2.25$ \\ 
$5.1\cdot 10^{19}$ &  $3.50$ \\
$1.0\cdot 10^{20}$ & $4.50$ \\
$2.0\cdot 10^{20}$ & $6.25$ \\
$4.1\cdot 10^{20}$ & $9.00$ \\
$1.0\cdot 10^{21}$ & $14.0$\\
\hline
\end{tabular}
\end{center}
\end{table}

\end{backmatter}
\end{document}